\pdfoutput=1
\documentclass[a4paper,10pt,twoside]{article}  
\usepackage{expl3,l3keys2e}      

\usepackage{type1ec}             

\usepackage[T1]{fontenc}         
\usepackage[utf8]{inputenc}      
\usepackage{mparhack,relsize}    
\usepackage{microtype}           
\usepackage{palatino}            
\usepackage[scaled=0.85]{beramono}  
\usepackage{textcomp}            
\usepackage[ngerman,english,main=italian]{babel}  

\usepackage{amscd,amsmath,amsfonts,amssymb,amsthm}     
\usepackage{bm}                  
\usepackage{braket}              
\usepackage{mathrsfs}            
\usepackage{mathtools}           
\usepackage{tensor}              
\usepackage[output-decimal-marker={.}]{siunitx} 
\usepackage[only,llbracket,rrbracket]{stmaryrd} 
\usepackage{upgreek}             

\usepackage{booktabs}            
\usepackage{caption}             
\usepackage{graphicx}          
\usepackage{bmpsize}             
\usepackage{float}               
\usepackage{multirow}            
\usepackage{rotating}            
\usepackage{subfig}              
\usepackage{tabularx}            
\usepackage[svgnames,dvipsnames,table]{xcolor}  
\usepackage[affil-it,auth-sc]{authblk}           

\usepackage{chngpage,calc,eso-pic,everyshi,ifthen} 
\usepackage{enumerate}
\usepackage[top=32mm,bottom=32mm,left=33mm,right=33mm,heightrounded,bindingoffset=5mm]{geometry}    
\usepackage{indentfirst}         
\usepackage{multicol}            
\usepackage[font=small,noorphans]{quoting} 
\usepackage{sectsty}             
\usepackage{setspace}            
\usepackage[nottoc,notlof,notlot]{tocbibind}  
\usepackage{titlesec}            

\usepackage{cite}
\usepackage{hyperref}            

\usepackage{bookmark}            

\hypersetup{
  breaklinks=true, pdfpagemode=UseNone, pageanchor=true,
  plainpages=false, bookmarksnumbered, bookmarksopen=true, bookmarksopenlevel=2, 
 	hypertexnames=true, pdfhighlight=/O,
  pdftitle={Deformation and dynamic response of abdominal aortic aneurysm sealing},
  pdfauthor={Argani},
  pdfsubject={Abdominal aortic aneurysm, endovascular sealing, migration, separation, simulations, fluid-structure interaction, contact mechanics},
  pdfkeywords={Abdominal aortic aneurysm, endovascular sealing, simulations, migration, separation, fluid-structure interaction, contact mechanics} }

\usepackage{fancyhdr}            
\fancypagestyle{plain}{%
  \fancyhead{}                                     
  \fancyfoot[CE,CO]{\bfseries\small\thepage}       
  \fancyhead[CE,CO]{ \textcolor{Maroon}{\footnotesize\sf{Abdominal aortic aneurysms and endovascular sealing: deformation and dynamic response}} }   
\setlength{\headheight}{14.6pt}}

\begin{document}
\selectlanguage{english}

\title{Abdominal aortic aneurysms and endovascular sealing: deformation and dynamic response}

\author[1]{L.P. Argani}
\author[2]{F. Torella}
\author[2]{R.K. Fisher}
\author[3]{R.G. McWilliams}
\author[4]{M.L. Wall}
\author[1]{A.B. Movchan}

\affil[1]{Department of Mathematical Sciences, University of Liverpool, UK}
\affil[2]{Liverpool Vascular \& Endovascular Service, Liverpool, UK}
\affil[3]{Royal Liverpool \& Broadgreen University Hospitals National Health Service Trust, Liverpool, UK}
\affil[4]{Russells Hall Hospital, The Dudley Group NHS Foundation Trust, Dudley, UK}

\date{}
\maketitle

\begin{abstract}
\noindent
Endovascular sealing is a new technique for the repair of abdominal aortic aneurysms.
Commercially available in Europe since~2013, it takes a revolutionary approach to aneurysm repair through minimally invasive techniques.
Although aneurysm sealing may be thought as more stable than conventional endovascular stent graft repairs, post-implantation movement of the endoprosthesis has been described, potentially leading to late complications.
The paper presents for the first time a model, which explains the nature of forces, in static and dynamic regimes, acting on sealed abdominal aortic aneurysms, with references to real case studies.
It is shown that elastic deformation of the aorta and of the endoprosthesis induced by static forces and vibrations during daily activities can potentially promote undesired movements of the endovascular sealing structure.
\end{abstract}

\section[Introduction]{Introduction}
  \label{Sezione-01}

The first attempt of an abdominal aortic aneurysm ligation was made back in~1817 by Sir Astley Cooper, and the first successful aneurysm repair was performed in~1951 by Charles Dubost.
The operation remained a great challenge for many years, and Albert Einstein was among those who died as a result of an abdominal aneurysm rupture in~1955.

\begin{figure}[tp]
\centering
\includegraphics[width=0.8\columnwidth,keepaspectratio]{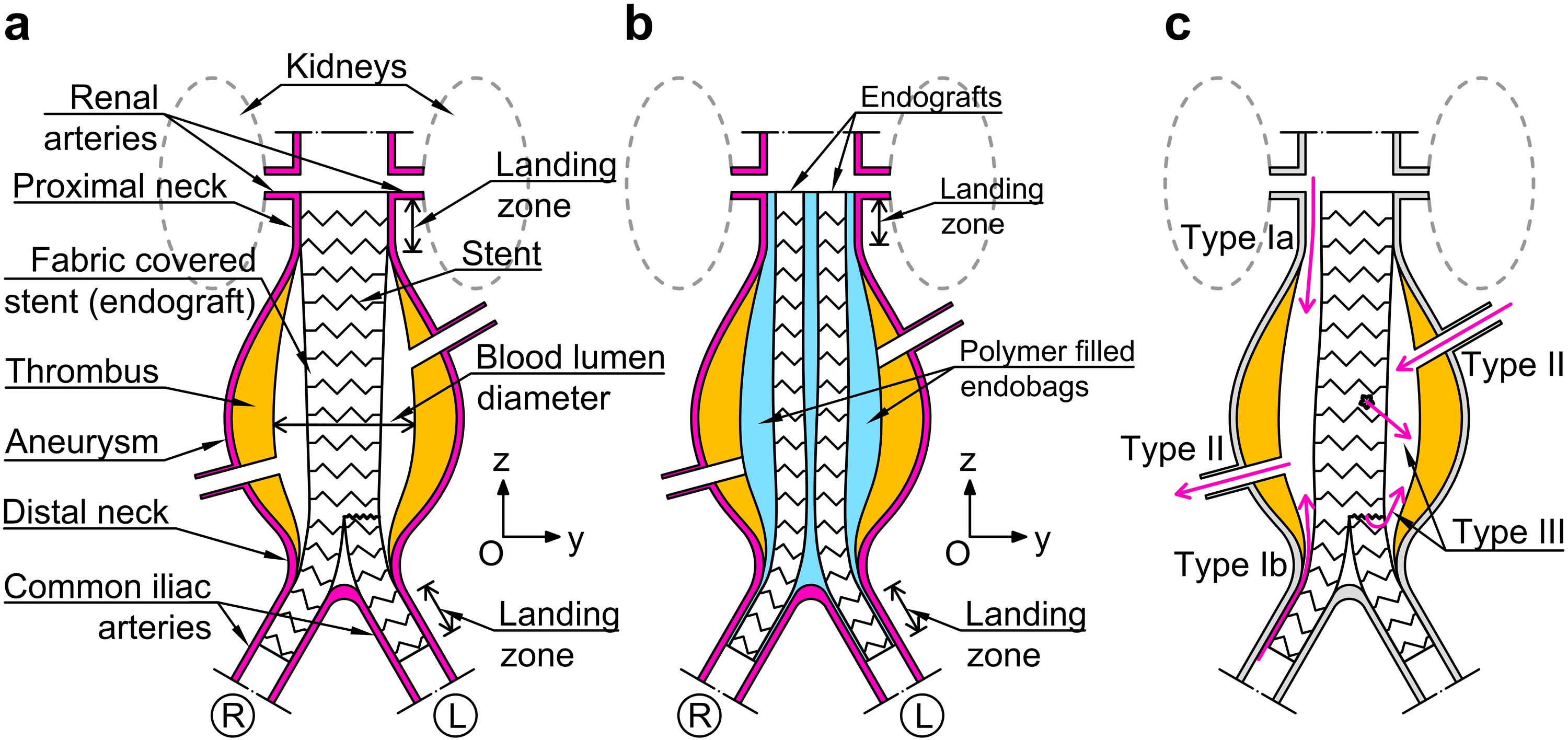}
\caption[Comparison between EVAR and EVAS sytems]{
Comparison between EVAR and EVAS sytems.
a, Schematic representation of the geometry of an AAA with an EVAR endograft; a three-dimensional reference system ($ Oxyz $) is introduced (the $x$-axis is perpendicular to the plane of the page), whereas circled letters~L and~R denote respectively the left and the right of the patient.
b, Reference scheme of the same AAA with the EVAS system.
c, Schematic representation of possible failure types of EVAR.
}
\label{fig:Figura-1}
\end{figure}

Aneurysms of the aorta are common, they most often involve the infra-renal segment, as shown in Fig.~\ref{fig:Figura-1}, and, in many cases, are treated by endovascular aneurysm repair (EVAR), which is less invasive than conventional surgery~\cite{Paravastu-Jayarajasingam-Cottam-Palfreyman-Michaels-Thomas:2014, Buck-vanHerwaarden-Schermerhorn-Moll:2014}.
Traditional aortic endografts are composed of tubes of thin surgical fabric reinforced with a metallic skeleton and are inserted through small groin incisions via the femoral arteries; endografts exclude abdominal aortic aneurysms from the circulation by anchoring themselves onto sections of normal artery proximal and distal to the dilated segment (the `landing zones') by means of barbs or hooks, and by exerting radial force on the arterial wall.
Despite their proven effectiveness, in a proportion of patients these endografts fail by allowing blood to re-enter the aneurysm.
This phenomenon is called `endoleak', which is classified depending on the causes of this blood flow; in particular, the main types of endoleaks are illustrated in Fig.~\ref{fig:Figura-1}c.
Type~Ia and~Ib correspond, respectively, to a leak around the proximal and the distal landing zones of the endoprosthesis; type~II corresponds to a persisting communication between one or more aortic side-branches and the aneurysm; type~III corresponds to a leak due to holes in the fabric cover or to disconnection between the endograft modules.~\cite{Antoniou-Georgiadis-Antoniou-Neequaye-Brennan-Torella-Vallabhaneni:2015}
The EVAR~1 trial suggested that the need for reintervention for endoleaks was approximately~25\% in the life of the study; reintervention is costly and time consuming for the health care professionals and for the patients and great benefit would be gained if it could be avoided~\cite{Patel-Sweeting-Powell-Greenhalgh:2016, The-United-Kingdom-EVAR-Trial-Investigators:2010}.

Endovascular sealing (EVAS) is a new technique to treat abdominal aortic aneurysms (AAA).
The concept underlying EVAS is original: the lumen of the AAA is completely filled by a bio-compatible polymer contained in endobags traversed by covered stents, which maintain perfusion of the iliac arteries.
Unlike traditional aortic endografts, the Nellix\textregistered{} (\textcopyright{} Endologix Inc. Irvine, Ca, USA) endoprosthesis, presently the only endograft available for EVAS, remains in place by virtue of the fact that, following its deployment, there is no room for displacement, thus eliminating any potential space for endoleaks.
EVAS may be a simpler, quicker procedure than EVAR, thus may reduce cost and be safer to the patient, and may result in lesser radiation use than EVAR, thus potentially less harmful for patient and the operating team~\cite{Antoniou-Senior-Iazzolino-England-McWilliams-Fisher-Torella:2016}.
Fig.~\ref{fig:Figura-1}a and Fig.~\ref{fig:Figura-1}b show the structural difference between EVAR and EVAS.

The EVAS concept assumes that the landing zones do not dilate in time, that the depressurised AAA does not change in size or shape, and that the thrombus often contained within AAAs remains unaltered in volume and distribution post-implantation, as any of these processes could result in endoleak or endograft displacement (`migration').
Clinical experience is limited but results of EVAS have been promising thus far~\cite{Carpenter-Cuff-Buckley-Healey-Hussain-Reijnen-Trani-Bockler:2016, Thompson-Heyligers-Hayes-Reijnen-Bockler-Schelzig-deVries-Krievins-Holden:2016, Bockler-Holden-Thompson-Hayes-Krievins-deVries-Reijnen:2015}, although migration of the Nellix\textregistered{} endoprosthesis has been described~\cite{England-Torella-Fisher-McWilliams:2016}, suggesting that one or more of the above assumptions may not hold true in every case.
It is unclear, however, why or how post-EVAS migration occurs: intuitively, the mechanisms leading to this complication will be very different than those observed following implantation of standard endografts, because the Nellix\textregistered{} endoprosthesis is radically different from conventional vascular stent grafts.
A significant difference from EVAR is that it has a greater mass dependent on the volume of the endobags polymer filler.
As the effects of static and dynamic loads on an object are highly dependent on its mass, it is likely that in vivo forces may have a radically different effect on patients treated with EVAS, as compared to those treated with standard endografts.
The aim of this research was to produce mathematical models to describe and quantify the effect of these forces on EVAS in vivo.

The new model of the EVAS systems incorporates fluid-solid interaction and contact interfaces.
The patient's immediate post-operative CT scan was used to build the geometry in the computational model, based on the novel mathematical approach.
The findings of the model were obtained before the clinical information about the evolution of the EVAS and subsequent observations in these case studies became available.

\section[Practical challenge and modelling]{Practical challenge and modelling}
  \label{Sezione-02}

A three-dimensional mathematical model of an AAA treated with EVAS was developed.
In particular, the model was focussed on the zone of the abdominal aorta between the renal arteries and the common iliac bifurcations, as sketched in Fig.~\ref{fig:Figura-1}b; these two branch points have been treated as the geometrical limit of the system, which includes the aortic wall, the thrombus layer, and two endografts enclosed into polymer-filled endobags.
The endografts have a sufficient length to cover the infrarenal aorta and most of the common iliac artery length.

Several mathematical models based on idealised and realistic geometries are provided in literature, which show the effects of the aortic bulging on the redistribution of the pressure and velocity of the blood and the subsequent stresses in the aortic wall, compared to a healthy artery~\cite{Boutsianis-Guala-Olgac-Wildermuth-Hoyer-Ventikos:2008, Figueroa-Taylor-Yeh-Chiou-Zarins:2009, Figueroa-Taylor-Yeh-Chiou-Gorrepati-Zarins:2010, Frauenfelder-Lotfey-Boehm-Wildermuth:2006, Georgakarakos-Xenakis-Georgiadis-Argyriou-Antoniou-Schoretsanitis-Lazarides:2014, Howell-Kim-Cheer-Dwyer-Saloner-Chuter:2007}; the effects of stent-graft EVAR on the interaction between the aorta and the blood flow have also been investigated in idealised and real case studies~\cite{Casciaro-Alfonso-Craiem-Alsac-ElBatti-Armentano:2016, Li-Kleinstreuer:2005b, Li-Kleinstreuer-Farber:2005, Li-Kleinstreuer:2006a, Scotti-Jimenez-Muluk-Finol:2008, DiMartino-Guadagni-Fumero-Ballerini-Spirito-Biglioli-Redaelli:2001, Molony-Callanan-Kavanagh-Walsh-McGloughlin:2009, Molony-Kavanagh-Madhavan-Walsh-McGloughlin:2010, Scotti-Finol:2007, Wolters-Rutten-Schurink-Kose-deHart-vanDeVosse:2005}.

\begin{figure}[tp]
\centering
\includegraphics[width=0.8\columnwidth,keepaspectratio]{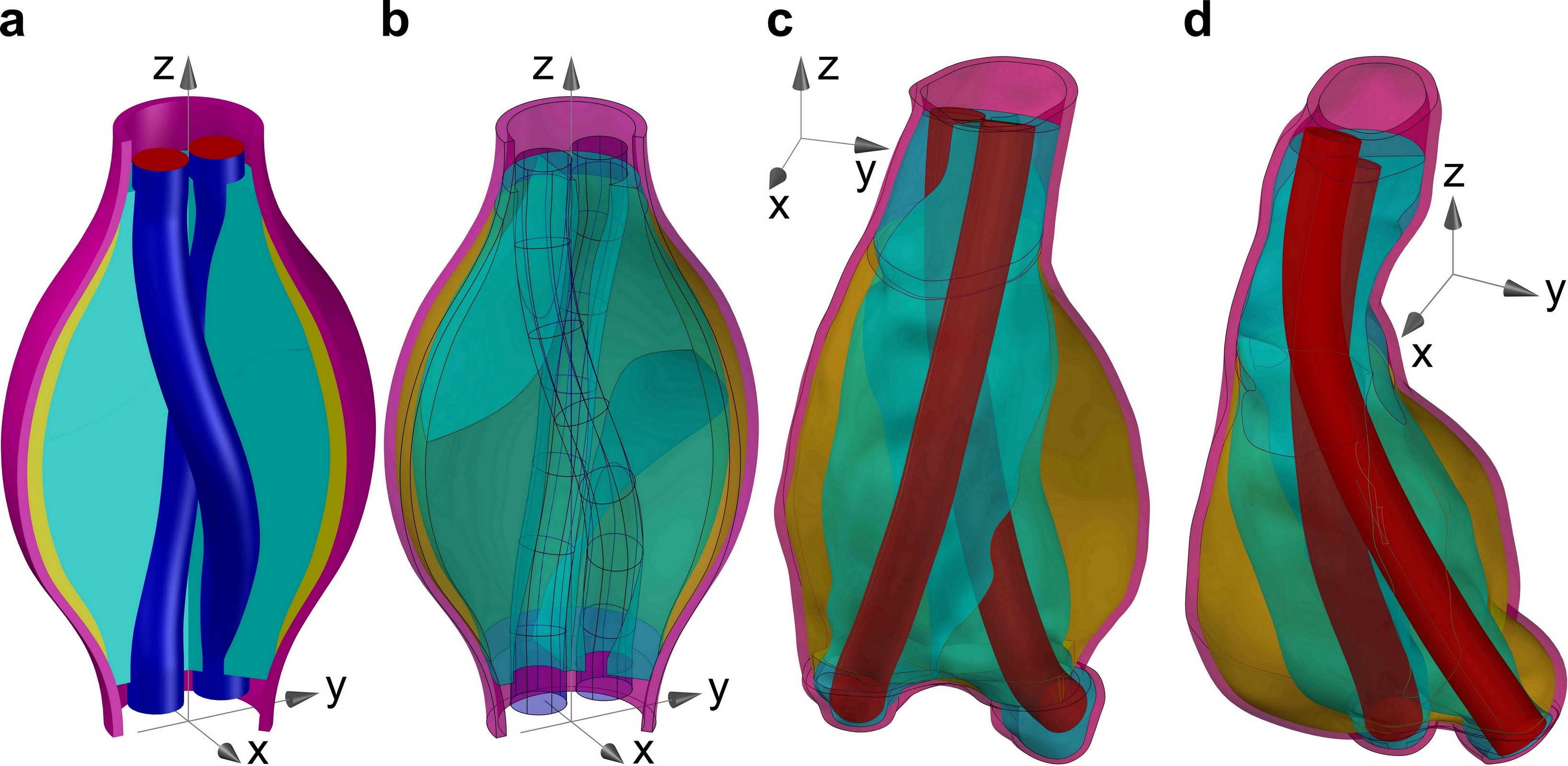}
\caption[Mathematical model geometry of post-EVAS AAA]{
Mathematical model geometry of post-EVAS AAA.
a, Solid cut-view of the idealised geometry of a sealed AAA with helical-like shape stents.
b, Transparent view of the idealised geometry of a sealed AAA showing the surface separating the two polymer filled endobags.
c, Transparent view of the sealed AAA for Patient 1.
d, Transparent view of the sealed AAA for Patient 2.
}
\label{fig:Figura-2}
\end{figure}

The aim of this work was to analyse elastic displacements and strains in a model of a sealed AAA, whose geometry was extrapolated from realistic medical measurements. 
wo cases of patients treated by EVAS.
The endoprosthesis had been placed in accordance with the indications for use prescribed by the manufacturer.
The drawings presented in Figs.~\ref{fig:Figura-2}c and~\ref{fig:Figura-2}d represent the anatomy on the basis of the post-operative CT scans.
The objective was to provide a predictive analysis of mechanisms, which may be responsible for migration and separation of components of the AAA sealing system in the postoperative period.
Furthermore, an idealised geometry was also considered in this work with the aim to show the role of boundary constraints, supporting clinical observations of EVAR reported in the literature~\cite{Heikkinen-Alsac-Arko-Metsanoja-Zvaigzne-Zarins:2006, Benharash-Lee-Abilez-Crabtree-Bloch-Zarins:2007, Resch-Malina-Lindblad-Malina-Brunkwall-Ivancev:2000} and connecting this concept to EVAS.
In particular, the papers~\cite{Heikkinen-Alsac-Arko-Metsanoja-Zvaigzne-Zarins:2006, Benharash-Lee-Abilez-Crabtree-Bloch-Zarins:2007} have addressed the role of distal fixations for EVAR systems, and the clinical results~\cite{Resch-Malina-Lindblad-Malina-Brunkwall-Ivancev:2000} are concerned with the importance of the proximal fixation in EVAR installations.
This idealised system is depicted in Figs.~\ref{fig:Figura-2}a and~\ref{fig:Figura-2}b, where the aorta and the thrombus have a circular cross-section only within the proximal and distal neck zones, whereas in the central part (the aneurysm sac) they have an elliptical cross-section.

The model includes two parts, which complement each other in the predictive analysis of the AAA repaired with EVAS: (1) the `modal analysis', which identifies the dynamic response of the system in absence of driving or damping forces in terms of natural (resonant) frequencies and their relative vibration modes, where all the components of the system are oscillating at the same frequency in absence of any driving or damping force; (2) the `static analysis', which accounts for mean static blood pressure and gravity force acting in different regimes, including the cases when a patient stands vertically, or is lying prone or sideways after the surgery.

The modal analysis illustrates the relative movements of the system that can be activated when a patient is subjected to vibrations.
Because this analysis takes into account the inertia of the entire system, special attention is given to the presence of the polymer filler contained in the endobags.
It can be noted that an additional mass associated with the polymer filler (\SI{2000}{\kilogram.\metre^{-3}} density) may reach up to~\num{500}-\SI{700}{\gram}, which would significantly lower the first natural frequency of the sealed AAA system, compared to the AAA without an additional mass.
The static analysis allows for the evaluation of the total displacement of the endoprosthesis during the immediate post-operative period; different configurations of the gravity force are representative of the positions assumed by a patient during daily life.

One of the most challenging features of the modelling proposed in this work, which differentiates it from the current literature on computational vascular biomechanics~\cite{Boutsianis-Guala-Olgac-Wildermuth-Hoyer-Ventikos:2008, Figueroa-Taylor-Yeh-Chiou-Zarins:2009, Figueroa-Taylor-Yeh-Chiou-Gorrepati-Zarins:2010, Frauenfelder-Lotfey-Boehm-Wildermuth:2006, Georgakarakos-Xenakis-Georgiadis-Argyriou-Antoniou-Schoretsanitis-Lazarides:2014, Howell-Kim-Cheer-Dwyer-Saloner-Chuter:2007, Casciaro-Alfonso-Craiem-Alsac-ElBatti-Armentano:2016, Li-Kleinstreuer:2005b, Li-Kleinstreuer-Farber:2005, Li-Kleinstreuer:2006a, Scotti-Jimenez-Muluk-Finol:2008, DiMartino-Guadagni-Fumero-Ballerini-Spirito-Biglioli-Redaelli:2001, Molony-Callanan-Kavanagh-Walsh-McGloughlin:2009, Molony-Kavanagh-Madhavan-Walsh-McGloughlin:2010, Scotti-Finol:2007, Wolters-Rutten-Schurink-Kose-deHart-vanDeVosse:2005}, is represented by the interaction between the components of the EVAS system.
In particular, the so-called contact interaction (namely, the analysis of the behaviour of solids touching each other at one or more points or zones) was implemented, taking into account the possibility of slipping (with and without friction) and separation of the EVAS components, thus allowing for the detection of possible relative displacement and/or the separation between the endobags as well as between the endobags and the thrombus/aorta.
Another important feature implemented was the mutual interaction between the blood and the structure within the analysis of the resonant frequencies of the system.

The purpose of the present paper is to introduce the new methodology of the study based on the analysis of partial differential equations describing a dynamic response of the EVAS system, which is typically neglected, rather than statistical analysis of a large dataset.
The frictional contact static problem solved in the context of EVAS, has been presented for the first time here to demonstrate possible slippage and separation of the EVAS components, which has been observed but not addressed in the past in any modelling work. 

\begin{figure}[tp]
\centering
\includegraphics[width=0.99\columnwidth,keepaspectratio]{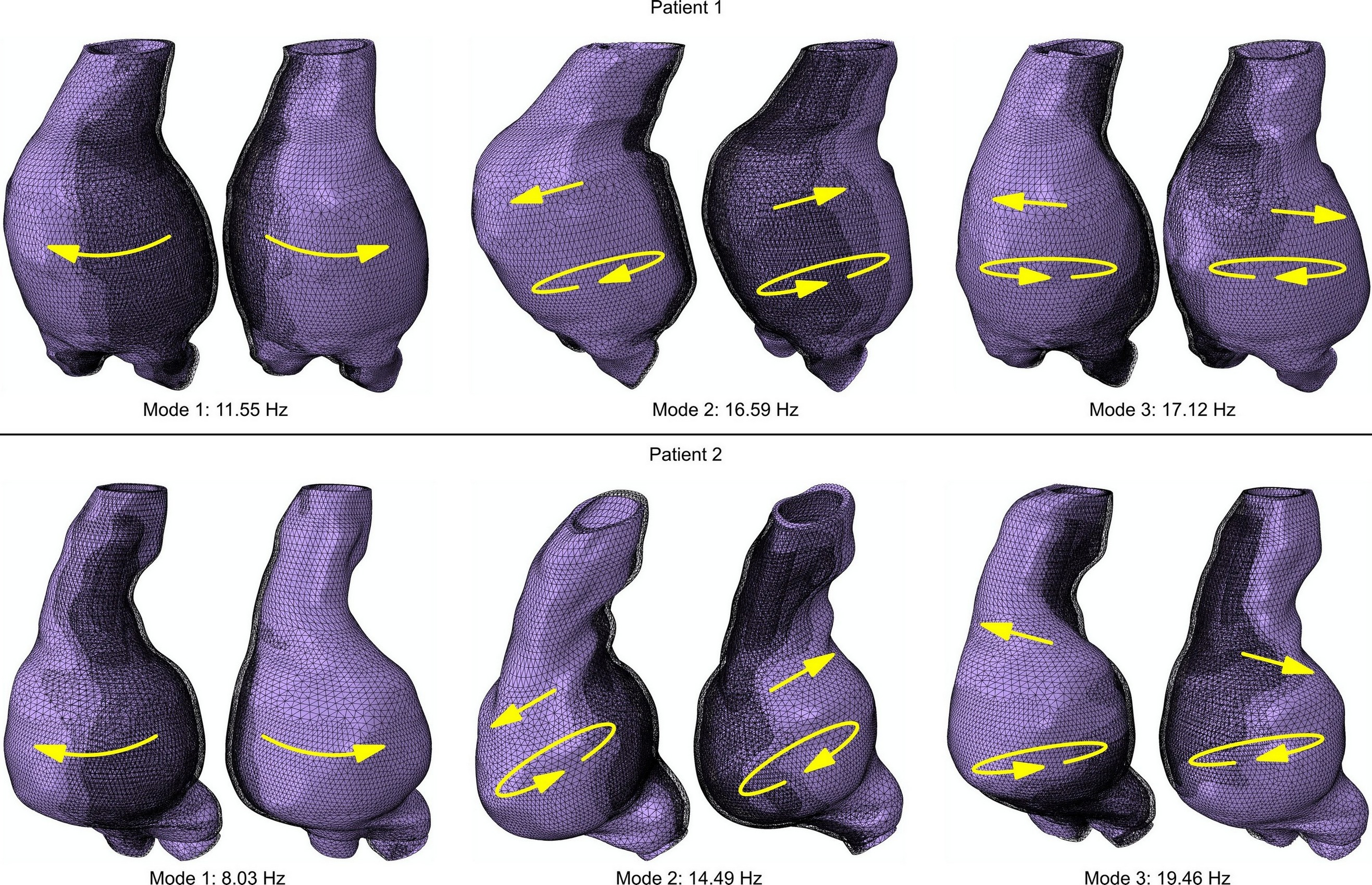}
\caption[First three vibration modes for the sealed aneurysm of two patients]{
First three vibration modes for the sealed aneurysm of two patients.
Deformed and undeformed shapes of the system are superimposed for each oscillatory motion (highlighted with arrows) for each vibration mode.
The higher the mode (and its relative frequency) the more complex is the shape; it can be noted that torsional-like movement affects the vibration modes for both patients already for the first three vibration modes.
The represented amplitude of vibration is intentionally magnified for clarity purposes.
}
\label{fig:Figura-3}
\end{figure}

\begin{table}[tp]
\caption[Resonant frequencies for the idealised model with different boundary conditions and for the patient specific models]{
Resonant frequencies for the idealised model with different boundary conditions and for the patient specific models.}
\label{tab:Tabella-1}
\centering
\begin{tabular}{cccccc}
\toprule
 \multirow{3}*{Mode}    & \multicolumn{3}{c}{Idealised model} & \multicolumn{2}{c}{\multirow{2}*{Patient specific model}} \\
 \cmidrule(lr){2-4}
   & \multicolumn{3}{c}{Endograft edges constrained}              &                       &                       \\
 \cmidrule(l){2-6}
   & None (free)        & Bottom             & Top and bottom     & Patient~1             & Patient~2             \\
\midrule
1  & \SI{4.32}{\hertz}  & \SI{8.58}{\hertz}  & \SI{10.86}{\hertz} & \SI{11.55}{\hertz}    & \SI{8.03}{\hertz}     \\
2  & \SI{10.15}{\hertz} & \SI{11.82}{\hertz} & \SI{15.91}{\hertz} & \SI{16.59}{\hertz}    & \SI{14.49}{\hertz}    \\
3  & \SI{10.22}{\hertz} & \SI{12.17}{\hertz} & \SI{16.37}{\hertz} & \SI{17.12}{\hertz}    & \SI{19.46}{\hertz}    \\
4  & \SI{16.37}{\hertz} & \SI{17.20}{\hertz} & \SI{17.32}{\hertz} & \SI{18.75}{\hertz}    & \SI{21.76}{\hertz}    \\
5  & \SI{17.09}{\hertz} & \SI{26.46}{\hertz} & \SI{33.81}{\hertz} & \SI{26.37}{\hertz}    & \SI{28.97}{\hertz}    \\
6  & \SI{18.07}{\hertz} & \SI{28.94}{\hertz} & \SI{34.62}{\hertz} & $ > \SI{32}{\hertz} $ & \SI{29.75}{\hertz}    \\
7  & \SI{21.15}{\hertz} & \SI{35.85}{\hertz} & \SI{42.83}{\hertz} & $ > \SI{32}{\hertz} $ & $ > \SI{30}{\hertz} $ \\
\bottomrule
\end{tabular}
\end{table}

\section[Modal analysis]{Modal analysis}
  \label{Sezione-03}

Every finite elastic system possesses resonant frequencies, which can be obtained by means of the modal analysis (eigenfrequencies).
If such frequencies are initiated by external means, then displacements associated with resonant vibrations may occur.
The nature of external forces, acting on a human body, may be linked to factors like walking, running, horse riding, using different means of transportation, using machinery which may produce vibrations.
If the first eigenfrequency is sufficiently high, this would imply that the patient is less likely to be affected by the low frequency dynamic input, which may result from everyday activities.
For instance, it shall be noted that the frequency that can be induced by running is approximately~\SI{3}{\hertz}, whereas public transportation (including the cases of rough street surface or speed humps) can easily induce vibrations in a range between~\SI{1}{\hertz} and~\SI{20}{\hertz} or wider, and hence falling within a range of interest for health and safety that is investigated by norms and other works about whole-body vibration exposure~\cite{Sekulic-Dedovic-Rusov-Salinic-Obradovic:2013, ISO-2631-1:1997, ISO-2631-4:2001}.

We shall use the term `frictional clamping' to identify constraints at the landing zones of the EVAS components.
This term is not conventional in medical practice, but it is important to acknowledge a constraint, which prevents the EVAS components from free vertical movement.
The numerical implementation is discussed in the Methods section.
The results of the modal analysis for the idealised model are summarised in Table~\ref{tab:Tabella-1}, showing clearly that a stronger (longer sealing zone) graft frictional clamping leads to higher resonant frequencies for the AAA repair system.
In particular, it can be noted that the first eigenfrequency can be relatively low (\SI{4.32}{\hertz}) in the case of a `weak' frictional clamping (i.e. when the endografts or the landing zones are too short).
In contrast, a `strong' distal frictional clamping (which is present if the bottom edges of the endografts are constrained) provides a higher first natural frequency (\SI{8.58}{\hertz}), which becomes even higher (\SI{10.86}{\hertz}) when strong proximal and distal frictional clampings (both top and bottom edges of the endografts are constrained) are implemented, so that in these cases resonant phenomena induced by vibrations are less likely to happen during a daily life routine.

The resonant frequencies obtained for the two patients are summarised in Table~\ref{tab:Tabella-1}, where it can be noted that the first resonant frequency (corresponding to a swinging sideways motion of the aneurysm sac) for Patient~1 (\SI{11.55}{\hertz}) is greater than the first resonant frequency for Patient~2 (\SI{8.03}{\hertz}), which makes the second patient potentially more vulnerable to the external dynamic forces.
Nevertheless, as in patients four resonant frequencies were in the range between~\SI{0}{\hertz} and~\SI{20}{\hertz}, both could still be subjected to undesired vibration induced movements of the endoprosthesis (migration and/or separation of the endobags) during routine activities.
The first three vibration modes and their corresponding frequencies for each of the two patients are depicted in Fig.~\ref{fig:Figura-3}.

\begin{figure}[tp]
\centering
\includegraphics[width=0.99\columnwidth,keepaspectratio]{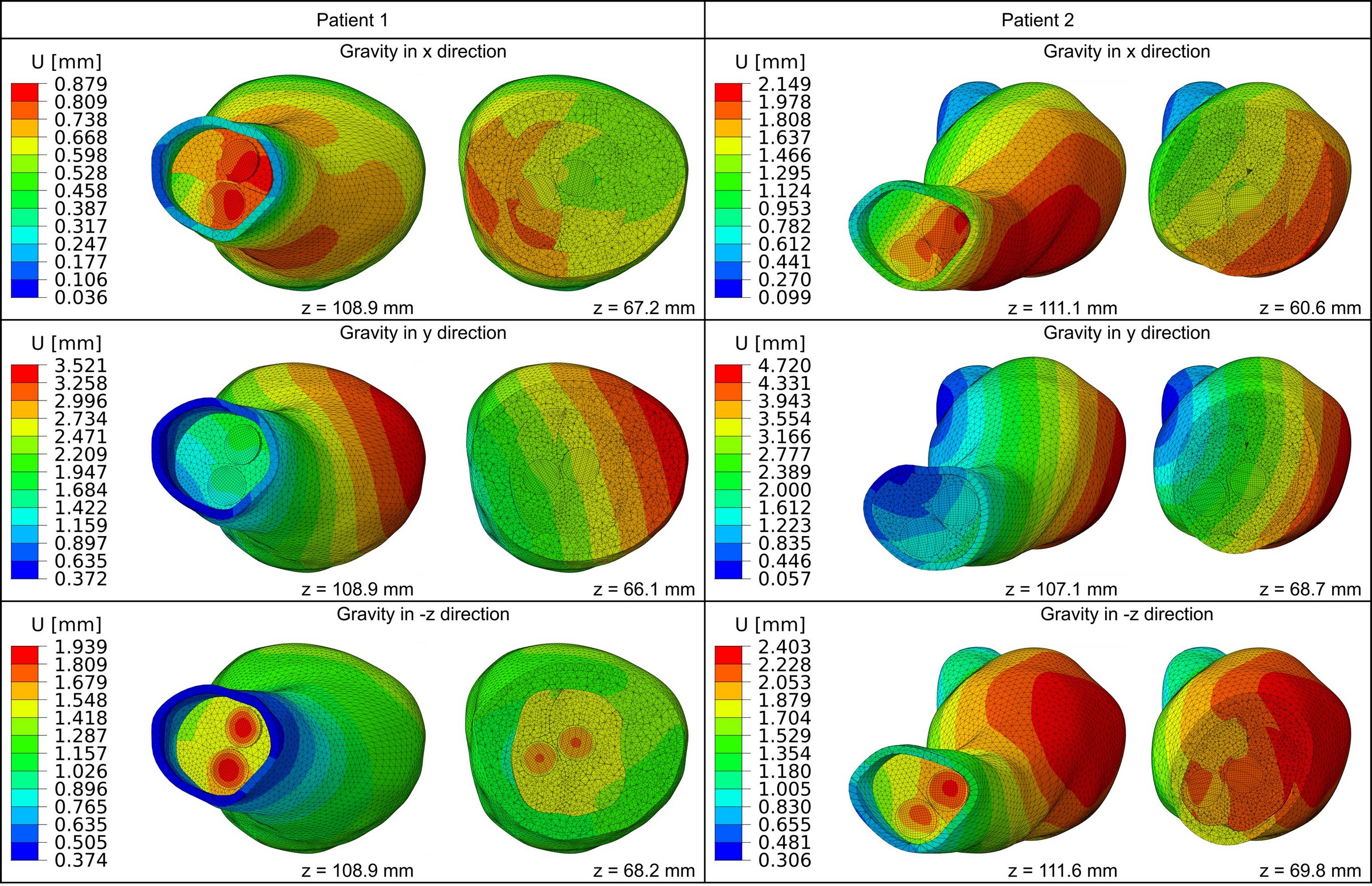}
\caption[Colour map of the displacement field magnitude U for different static load cases]{
Colour map of the displacement field magnitude U for different static load cases.
Discontinuities (jumps) in the displacement field may occur only across the components interfaces (contact surfaces) and denote relative movements between the components of the system; the endobags can slide together (bottom left) with respect to the aortic wall or separately (top left and bottom right).
The maps of the displacement magnitude are referred to the top and to the middle cross-sections of the aneurysm sac for both patients.
}
\label{fig:Figura-4}
\end{figure}

\begin{figure}[tp]
\centering
\includegraphics[width=0.99\columnwidth,keepaspectratio]{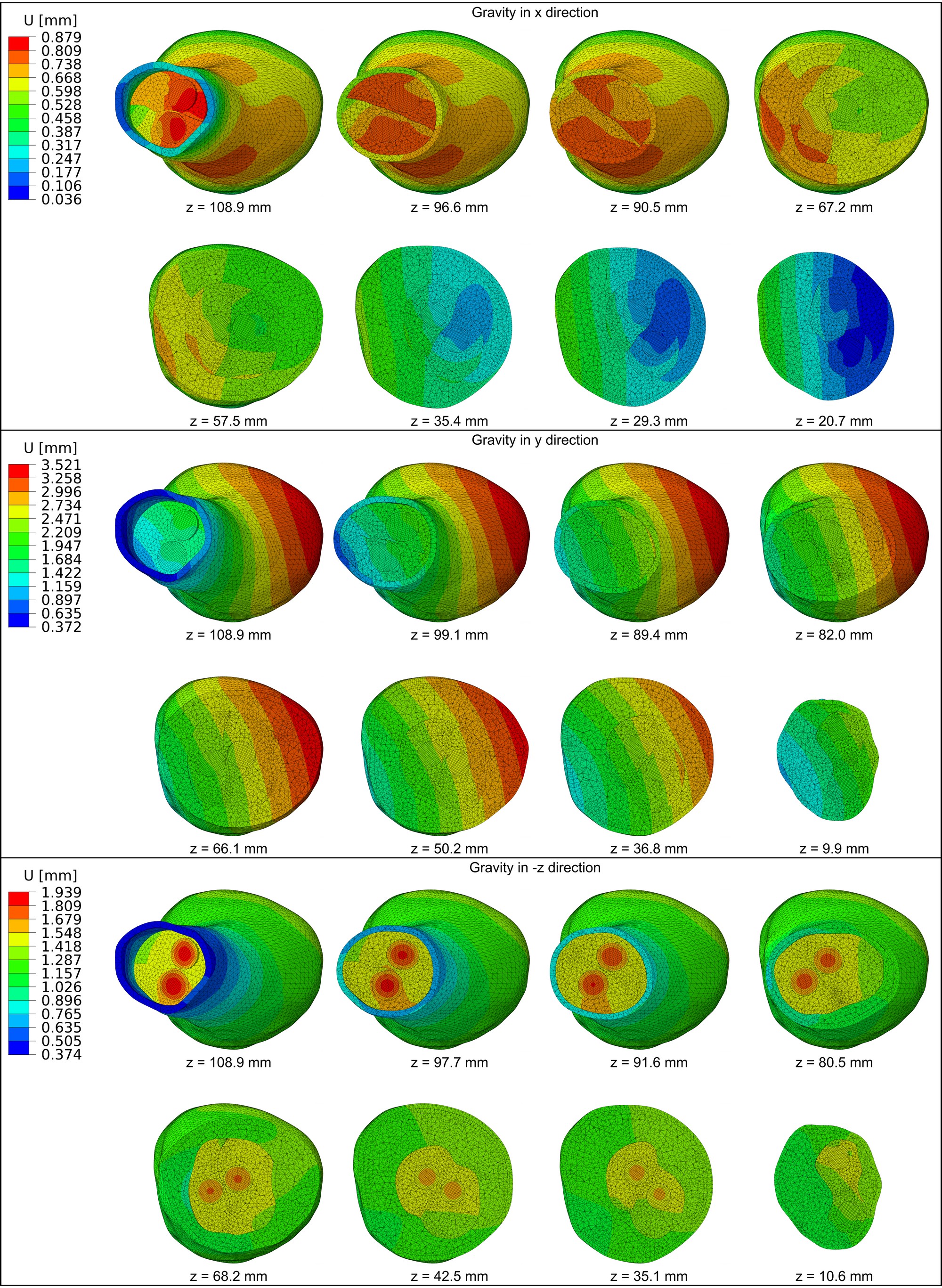}
\caption[Colour map of the displacement field magnitude for Patient~1]{Colour map of the displacement field magnitude $ U $ for Patient~1.
Discontinuities (jumps) in the displacement field may occur only across the components interfaces (contact surfaces) and denote relative movements between the components of the system.
}
\label{fig:Figura-S4}
\end{figure}

\begin{figure}[tp]
\centering
\includegraphics[width=0.99\columnwidth,keepaspectratio]{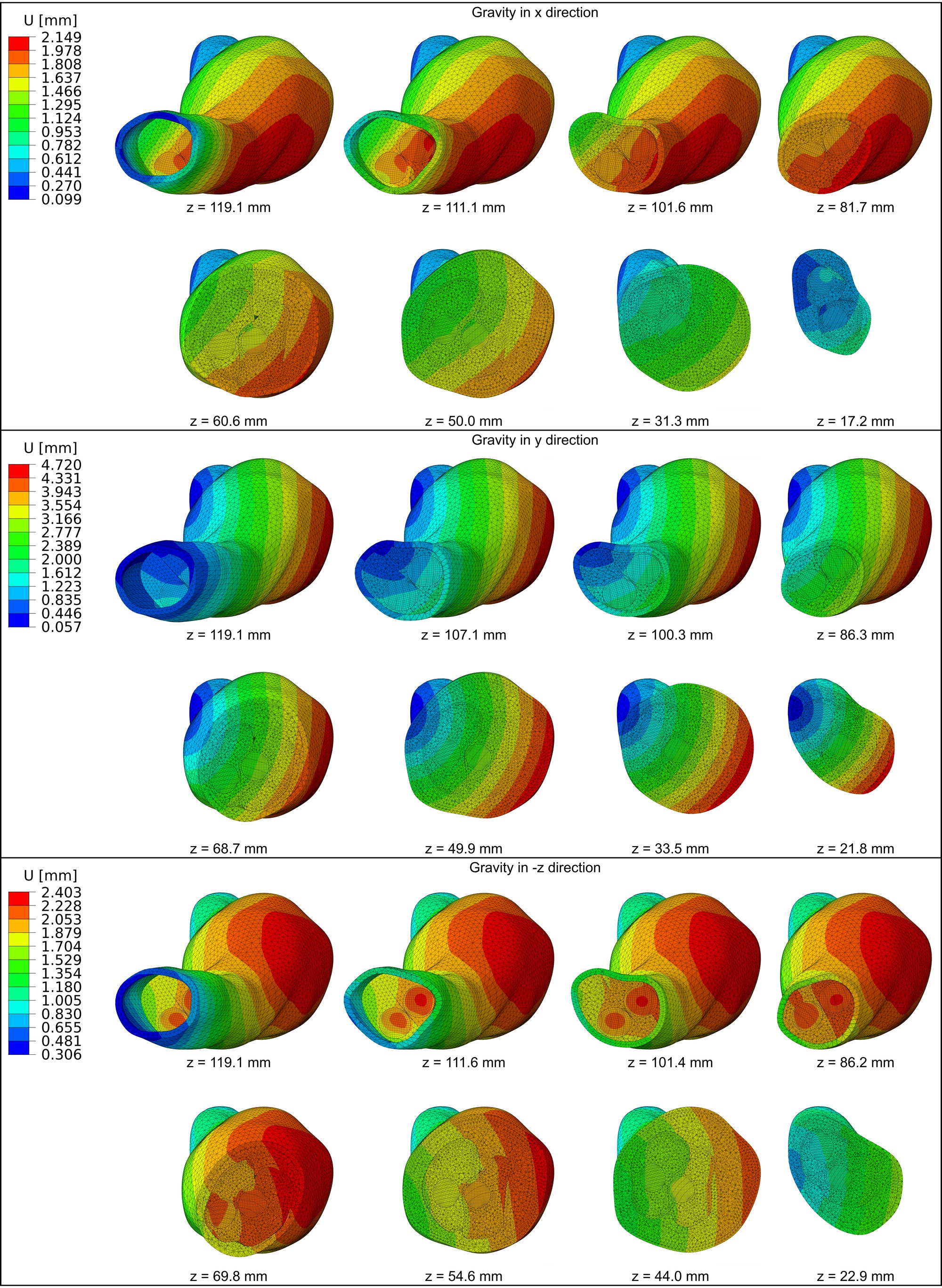}
\caption[Colour map of the displacement field magnitude for Patient~2]{Colour map of the displacement field magnitude $ U $ for Patient~2.
Discontinuities (jumps) in the displacement field may occur only across the components interfaces (contact surfaces) and denote relative movements between the components of the system.
}
\label{fig:Figura-S5}
\end{figure}

\section[Static forces and deformations]{Static forces and deformations}
  \label{Sezione-04}

Three configurations of static loading were investigated in the framework of the static analysis, corresponding to different directions of gravity (in particular the three principal directions of the three-dimensional reference system) with constant mean blood pressure.
Numerical simulations produced for both patients for aforementioned loadings identified displacement jumps across the interfaces of the EVAS components, as summarised in Fig.~\ref{fig:Figura-4}.
These include the interface between the endobags, as well as the interface between individual endobags and the thrombus/aorta; these jumps in displacement field denote relative movement between the EVAS components, thus demonstrating the possibility of migration.

\begin{figure}[tp]
\centering
\includegraphics[width=0.8\columnwidth,keepaspectratio]{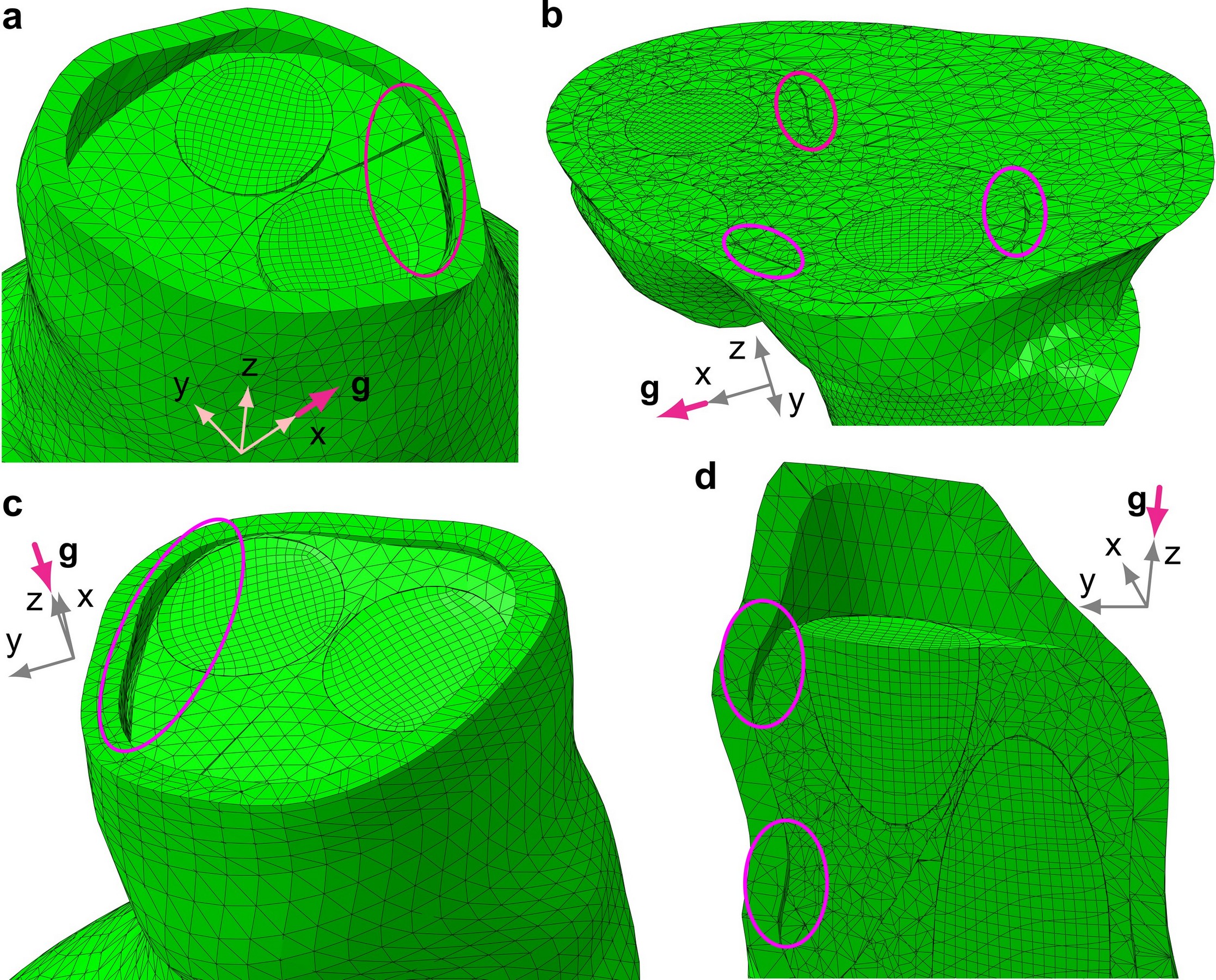}
\caption[Partial detachment of the endoprosthesis from the surrounding aortic and thrombus inner walls for different positions of a patient]{
Partial detachment of the endoprosthesis from the surrounding aortic and thrombus inner walls for different positions of a patient ($ \bm{g} $ is gravity).
Partial separation can occur simultaneously in interface zones within the sealed aneurysm.
a, Small void opening between the two endobags and the aorta (Patient~1).
b, Partial separation of the endoprosthesis from the intraluminal thrombus in interface zones within the aneurysm sac (Patient~1).
c, Small void opening at the top left interface zone (Patient~2).
d, Vertical cut-view of the partial detachment of the endoprosthesis from the upper aortic wall (Patient~2).
}
\label{fig:Figura-5}
\end{figure}

Static analysis also demonstrates that opening of small voids between the endobags may already occur in the immediate postoperative period.
Furthermore, small void opening can also occur simultaneously at many points between the endobags and the thrombus/aorta within the aneurysm sac and/or at the top of the landing zone, as depicted in Fig.~\ref{fig:Figura-5}.
It can be noted that partial detachment of the endoprosthesis from the thrombus and aortic walls leads to localised contact pressure, which increases the stresses on the aorta and, potentially, can promote thrombus displacement.

It is shown that gravity yields displacement of different magnitude depending on the direction of the force and that relative movements of the endoprosthesis are decreasing in the distal neck.
Detailed results are provided in Figs~\ref{fig:Figura-S4} and~\ref{fig:Figura-S5}.

The theoretical analysis suggests the possibility of separation of the endobags for Patient~2, where the first eigenfrequency was low.
The vibration modes as well as static deformations (for each force direction) showed space opening between the endobags, as well as between the endobags and the aorta in the proximal part of the sealed aneurysm (maximum depth and width detected equal to~\SI{6.71}{\milli\metre} and~\SI{1.02}{\milli\metre} respectively, see Fig.~\ref{fig:Figura-5}).
On the contrary, in Patient~1 the sealed AAA has a higher first eigenfrequency and relative displacements of endobags indicating a possible relative slippage rather than separation.
The numerical modelling presented here is deterministic and does not take into account the whole transient history of dynamic forcing.
It is used to predict the tendency to migration/separation of EVAS components rather than results of a long-time evolution.

The static analysis for the idealised model confirmed that graft clamping is important also for EVAS because the jumps in displacement field detected in the whole AAA repair system were considerable in the case of weak frictional clamping and almost negligible when double (top and bottom constraints) `strong' frictional clamping was provided.

The problem in hand has been solved directly through the partial differential equations analysis of boundary value problems and spectral problems.
The numerical results and theoretical concepts presented here are deterministic and do not involve a transient loading.
Although the computational model does not allow for a variety of probabilistic scenarios of possible dynamic transient loads, it provides a valuable insight on EVAS linking the strength of proximal frictional clamping and migration of the endoprosthesis under static and dynamic forces.
It is also noted that the dimensions of the endobags depend on the longitudinal and lateral sizes of the aneurysm and ideally the endobags filling occupies the entire aneurysm sac – otherwise relative movement of EVAS components may occur.
EVAS technique is designed in such a way that it does not exert radial forces on the landing zones during its deployment.
However, it is shown in the numerical simulations that contact pressure acting on proximal and distal necks, arising from static and dynamic loading, provides elastic deformation (and subsequent induced stresses) of these landing zones.

\section[From EVAR to dynamic EVAS]{From EVAR to dynamic EVAS}
  \label{Sezione-05}

Despite its innovative design, the Nellix\textregistered{} endoprosthesis may be more vulnerable to certain forces because of its mass.
The present study demonstrates that resonance may occur when the endograft is exposed to frequencies encountered during day-to-day activities, causing pendular movements which, in time, may contribute to endograft instability and migration.
Interestingly, fixation of the proximal and distal end of the endograft reduces the range of these frequencies, pushing the lowest towards the higher end of the spectrum experienced during daily activities.
Even gravity, modelled as a static load, may, potentially, result in displacement of the endobags that could allow blood to seep between them, or between the endobags and the aortic wall.
This may, in turn, create a wedge that in time could foster endobag separation and re-pressurization of the aneurysm.

The modelling and numerical simulations were performed for accurately reproduced geometry and physical conditions describing the sealed AAA.
The combination of the modal and static analysis suggests that the first patient may be prone to migration, and that, in the second case, separation of the endobags may occur.
Actual observations, three years after EVAS, have shown that, in the first patient, there has been progressive migration of~\SI{1}{\centi\metre} of the endobags, together with a small increase of AAA volume (maximum diameter increased~\SI{2}{\milli\metre}) without endoleak; in the second patient there has not been any migration nor endoleak but a relative separation (of~\SI{6}{\milli\metre}) of the EVAS components and a small increase of AAA volume have been detected (maximum diameter increased~\SI{3}{\milli\metre}).

The results of this theoretical study can be used in three ways.
First, they help to elucidate potential mechanisms of failure of the Nellix\textregistered{} endograft and, more generally, of the aneurysm sealing concept.
Second, as the effects of gravity and vibration appear dependent on the physical property of individual AAAs, these studies may be used in selection of patients whose aortic anatomy makes them more or less susceptible to the effect of these forces post-implant, ultimately enabling clinicians to recommend more personalised treatment strategies.
Third, they may inform future product designs, as manufacturers strive to produce new generations of endografts that are less susceptible to the effects of forces experienced by patients during their daily activities.

\section[Methods]{Methods}
  \label{Sezione-M-01}

\paragraph{Ethics statement.}
The work did not involve any experiments on humans; the work is purely theoretical and did not involve any experiment on human tissues.
The paper does not include any photographs, X-rays or CT scans of human organs.
The EVAS endoprosthesis, discussed in the paper, had been placed in accordance with the indications for use prescribed by the manufacturer.

\paragraph{Geometry for real patients.}
The geometry was imported from two sets of data for anonymised patients treated with EVAS in~2013 at Royal Liverpool and Broadgreen University Hospital.
For Patient~1 the AAA length and its maximal diameter are~\SI{107.3}{\milli\metre} and \SI{67.4}{\milli\metre} respectively, whereas for Patient~2 the length and the maximal diameter are~\SI{110.7}{\milli\metre} and~\SI{61.4}{\milli\metre} respectively.
The proximal neck length and angulation for Patient~1 are equal to~\SI{20}{\milli\metre} and~\SI{0}{\degree} respectively, while for Patient~2 they are equal to~\SI{45}{\milli\metre} and~\SI{5}{\degree} respectively.
In particular, the diameter of the top and of the bottom cross-sections of the proximal neck of Patient~1 is equal to~\SI{28}{\milli\metre} and~\SI{29}{\milli\metre} respectively, whereas the proximal neck of Patient~2 presents a constant cross-section with~\SI{24}{\milli\metre} diameter.
The principal diameters of the aortic bifurcation are equal to $ 17 \times \SI{27}{\milli\metre} $ for Patient~1 and $ 18 \times \SI{26}{\milli\metre} $ for Patient~2.
The maximum blood lumen diameter is equal to~\SI{38}{\milli\metre} for Patient~1, whereas for Patient~2 it is equal to~\SI{34}{\milli\metre}.

\begin{figure}[tp]
\centering
\includegraphics[width=0.9\columnwidth,keepaspectratio]{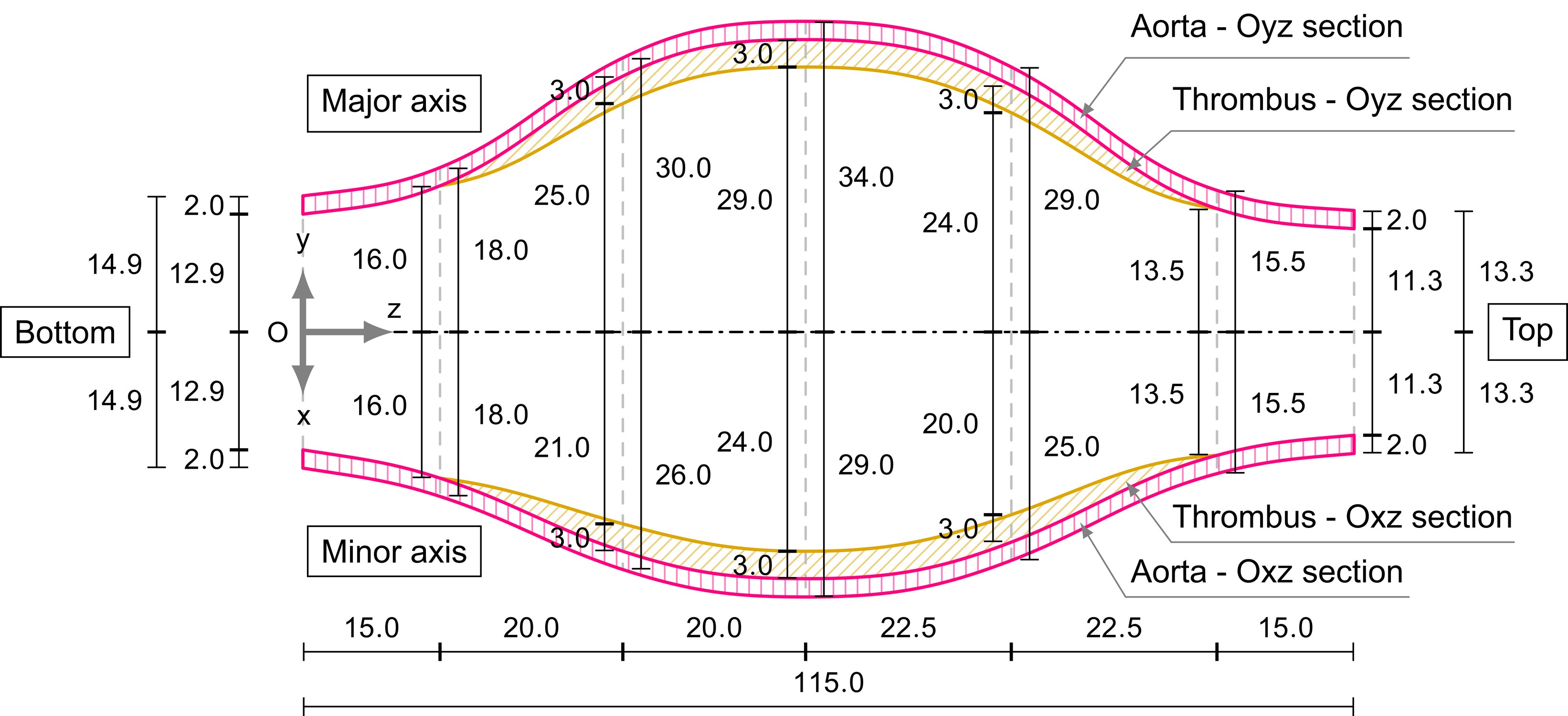}
\caption[Dimensions of the idealised model]{Dimensions of the idealised model.
All the measures reported are in millimetres.
}
\label{fig:Figura-S1}
\end{figure}

\paragraph{Idealised model geometry.}
The idealised model of a sealed AAA is based on a geometry with two symmetry axes ($ x $ and $ y $, $ z $ being the vertical axis); the main measures of the idealised model are summarised in Fig.~\ref{fig:Figura-S1}, from which it can be noted that the lengths of AAA is~\SI{115.0}{\milli\metre} whereas the maximal diameter of the bulged cross-section is~\SI{70}{\milli\metre}.
The aorta and the thrombus have a circular cross-section only within the proximal and distal neck zones, whereas in the central part (the aneurysm sac) they have an elliptical cross-section (semi-axes aligned with $ x $ and $ y $ directions).
The endografts have a circular cross-section (\SI{10}{\milli\metre} radius, \SI{0.1}{\milli\metre} thickness) and they are displaced in a helical-like shape, twisted by~\SI{180}{\degree} within the aneurysm sac, hence they do not have any symmetry; only their top and bottom sections are symmetric with respect to the $ x $ axis.
The polymer matrix is not symmetric because of the endograft shape and the subsequent twisted interface surface separating the two endobags; only the two separation lines between the endobags that are visible at the top and at the bottom of the endoprosthesis are aligned with the $ x $ axis, and hence these are its two unique symmetric points.

\paragraph{Forces and displacements in the mathematical model.}
The displacement field $ \bm{u}_{\textup{s}} $ characterises the motion of the elastic solid components of the AAA repair components, whereas the velocity field $ \bm{v}_{\textup{s}} $ corresponds to the fluid flow, with the fluid-solid interaction being taken into account.
Here the notations $ \mu_{\textup{f}} $, $ \rho_{\textup{f}} $ and $ p_{\textup{f}} $ are used for the viscosity, mass density and pressure within the fluid; $ \lambda $ and $ \mu $ are the Lamé elastic moduli of the solid phase; $ \bm{f} $ is the body force density, and $ \bm{\sigma}_{\textup{s}} $ is the stress tensor; the vector $ \bm{n} $ represents a unit normal on the solid/fluid interface.
The governing linearised equations describing the motion of the AAA fluid-solid system have the form
  \begin{subequations}
    \label{eq:moto}
      \begin{gather}
        \nabla \cdot \bm{v}_{\textup{f}} = 0   \label{eq:moto-continuita} \\
        \frac{ \partial \bm{v}_{\textup{f}} }{ \partial t } + \frac{ \nabla p_{\textup{f}} }{ \rho_{\textup{f}} }
        - \frac{ \mu_{\textup{f}} }{ \rho_{\textup{f}} } \, \nabla^{2} \bm{v}_{\textup{f}} = \bm{f}   \label{eq:moto-navier-stokes} \\
        \mu \, \nabla^{2} \bm{u}_{\textup{s}} + ( \lambda + \mu ) \nabla ( \nabla \cdot \bm{u}_{\textup{s}} )
        = \rho_{\textup{s}} \frac{ \partial^{2} \bm{u}_{\textup{s}} }{ \partial t^{2} }               \label{eq:moto-problema-elastico}
      \end{gather}
  \end{subequations}
accompanied by the interface conditions describing the fluid-solid interaction
  \begin{subequations}
    \label{eq:accoppiamento}
      \begin{gather}
        \bm{v}_{\textup{f}} = \frac{ \partial \bm{u}_{\textup{s}} }{ \partial t }  \label{eq:accoppiamento-velocita-spostamento} \\
        \bm{\sigma}_{\textup{s}} \, \bm{n} = - p_{\textup{f}} \, \bm{n}            \label{eq:accoppiamento-pressione-sforzi}
      \end{gather}
  \end{subequations}

\begin{figure}[tp]
\centering
\includegraphics[width=0.22\columnwidth,keepaspectratio]{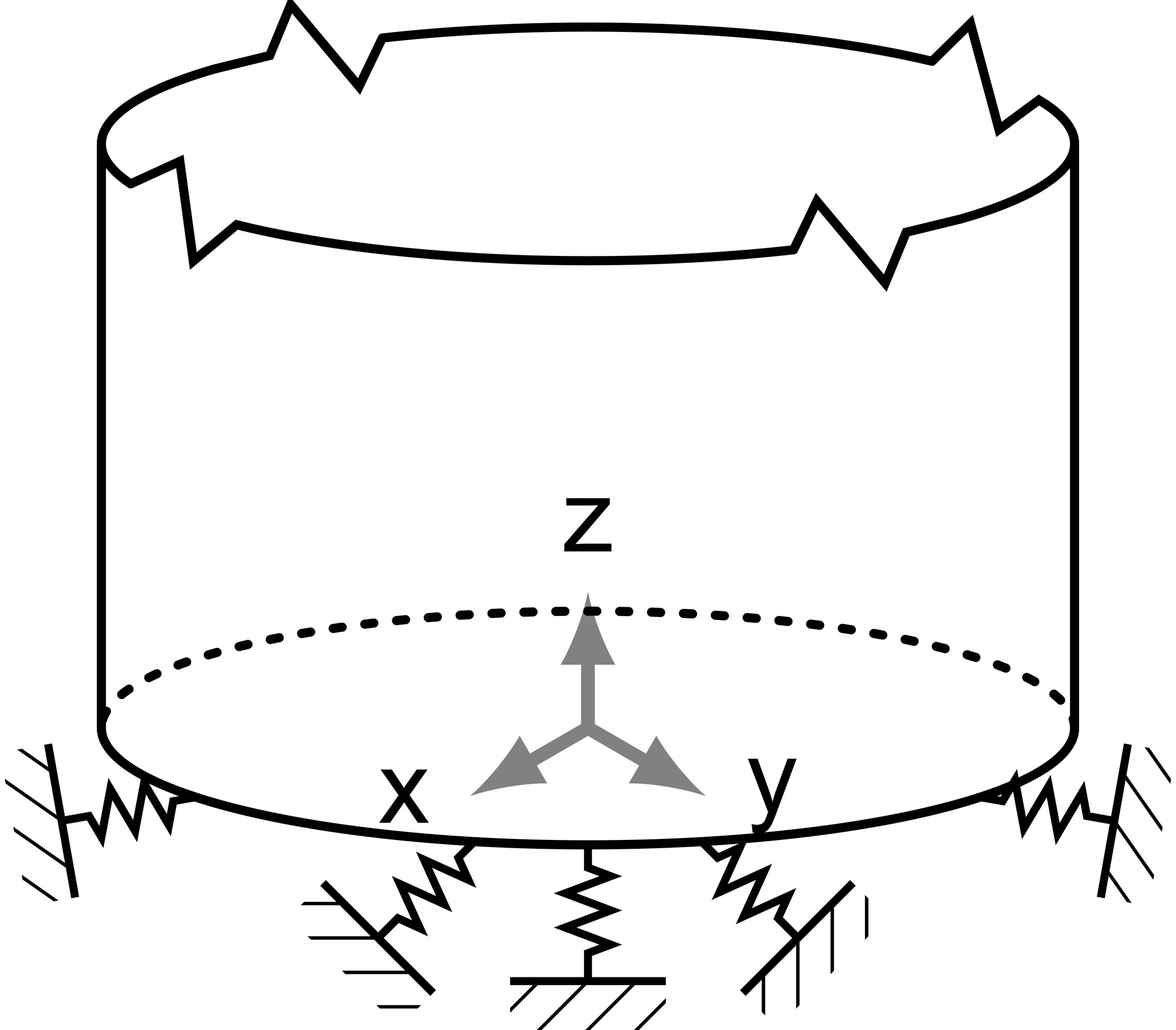}
\caption[Example of the spring-type boundary conditions]{Example of the spring-type boundary conditions.
This scheme represents the end section of a generic component of the aorta-endoprosthesis sealing system subjected to continuity conditions.
}
\label{fig:Figura-S2}
\end{figure}

\begin{figure}[tp]
\centering
\subfloat[][\emph{Example of uniaxial tensile test.} \label{fig:Figura-S3-a}]{
\includegraphics[width=0.24\columnwidth,keepaspectratio]{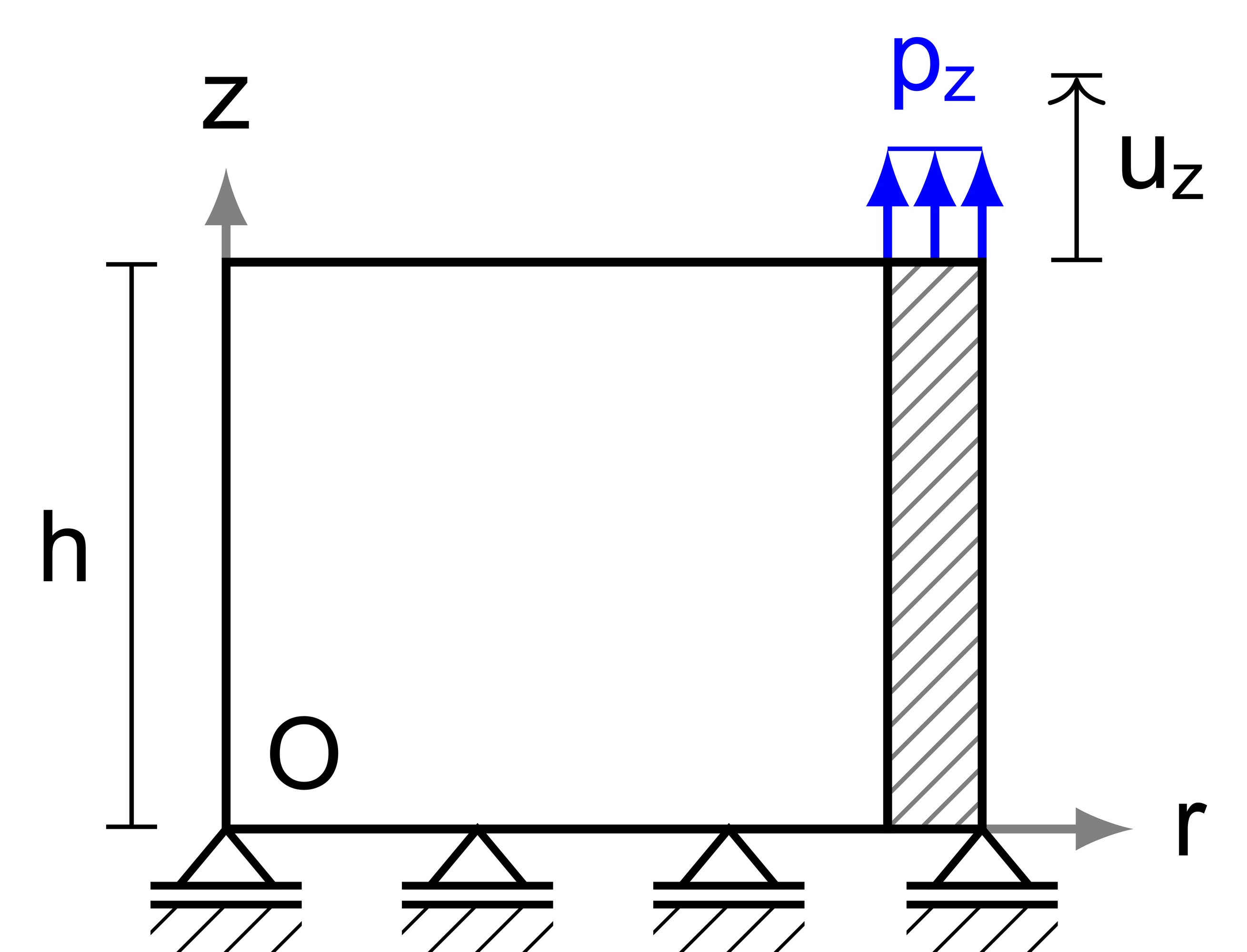}
}
\qquad
\subfloat[][\emph{Example of shear test.} \label{fig:Figura-S3-b}]{
\includegraphics[width=0.35\columnwidth,keepaspectratio]{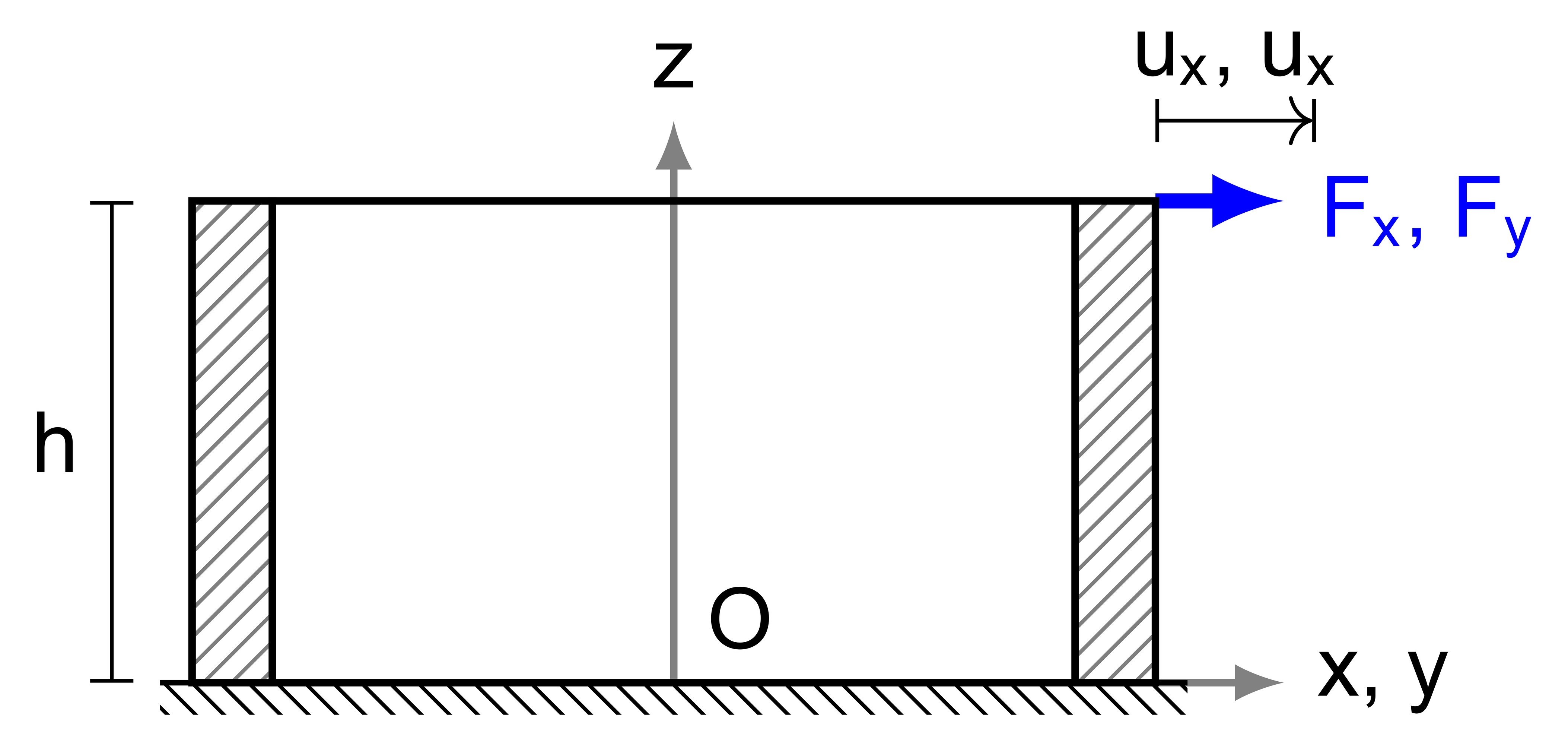}
}
\caption[Supplemental model for the calibration of the spring-type boundary conditions]{Supplemental model for the calibration of the spring-type boundary conditions.
Example of uniaxial tensile test~\subref{fig:Figura-S3-a} and shear test~\subref{fig:Figura-S3-b} on a tubular structure having the same cross section of the end section of an EVAS component subjected to continuity conditions.
}
\label{fig:Figura-S3}
\end{figure}

\paragraph{Boundary conditions for structural components and continuity condition.}
At the top and the bottom regions of the sealed AAA structure, the endografts are securely inserted into the artery, which is modelled by spring-type boundary conditions sketched in Fig.~\ref{fig:Figura-S2}.
Springs acting along the three principal directions are attached to the end sections of the aorta and of the endoprosthesis, thus providing the continuity condition of these parts.
These spring sets were calibrated through supplemental models, i.e. uniaxial tensile and shear tests, as depicted in Fig.~\ref{fig:Figura-S3}.
In the idealised model three limit cases were considered: (1) graft edges free of tractions (2) graft edges constrained at the bottom by springs, corresponding to the continuity condition that represents most common situation for patients; (3) graft edges constrained contemporarily at their top and bottom, an extension of the second case to the limit case of a long proximal landing zone (and long endografts) or a proximal fixation with barbs or hooks.

\paragraph{Contact interaction formulation.}
Frictional contact between the two endobags included in the EVAS system is allowed, as well as frictionless contact between the endobags and the thrombus/aorta.
This implies the continuity of the normal elastic displacement across the frictional interface and a constraint on the tangential traction along the interface: for the frictionless contact, the tangential traction is zero, whereas for the case of Coulomb's friction the tangential traction is proportional to the normal traction, with the proportionality factor being the friction coefficient of the value~\num{0.1}.
A particular feature of the condition of frictional (or frictionless) contact is the possibility of slippage along the interface boundary, which implies the discontinuity of the tangential displacement of solids in contact.
Zero tractions are set on the free surface, resulting from the separation of the EVAS components.

\paragraph{Material parameters.}
The coefficients in the equations~\eqref{eq:moto} require the knowledge of the Young's modulus, Poisson's ratio and the mass density for the components of the AAA sealing system.
These coefficients are summarised in Table~\ref{tab:Tabella-2}.
For the frequency analysis blood was modelled as an inviscid incompressible fluid with bulk modulus and density equal to~\SI{2.0}{\giga\pascal} and~\SI{1050}{\kilogram.\metre^{-3}} respectively, whereas for the static analysis a soft material with~\SI{0.9}{\kilo\pascal} Young's modulus and~\num{0.495} Poisson's ratio was employed.
The endobags are constituted by polyethylene glycol-based polymer (PEG-DA)~\cite{Gaebler-Stampfl-Seidler-Schueller-Redl-Juras-Trattnig-Weidisch:2009} whereas the endografts are constituted by expanded polytetrafluoroethylene (ePTFE).
Thrombus layer mean elastic properties from Computational Hemodynamic literature were employed, which are representative of the material behaviour for the analyses performed in this work.
The same material properties were employed in idealised and realistic models because no patient-specific data were available.

\begin{table}[tp]
\centering
\caption[Elastic properties of the materials]{Definition of the elastic properties of the materials employed in the mathematical model.}
\label{tab:Tabella-2}
\begin{tabular}{ccccc}
\toprule
\multirow{2}*{Mechanical properties} & \multicolumn{4}{c}{Components of the AAA sealing system} \\
\cmidrule(l){2-5}
                                     & Aorta      & Thrombus   & Endobag    & Graft      \\
\midrule
Young's modulus~[\si{\mega\pascal}]  & \num{0.8}  & \num{0.4}  & \num{0.09} & \num{500}  \\
Poisson's ratio                      & \num{0.49} & \num{0.45} & \num{0.49} & \num{0.46} \\
Density~[\si{\kilogram.\metre^{-3}}] & \num{1200} & \num{910}  & \num{2000} & \num{2200} \\
\bottomrule
\end{tabular}
\end{table}

\paragraph{Computational approach for modal and static analysis.}
The model was programmed into Abaqus Unified FEA (\textcopyright{} Dassault Systemes) as follows.
Hexahedral and tetrahedral shaped finite elements were employed to model, respectively, the regular and non-regular solid parts (i.e. regions including corners, wedges and cusps on the boundary, and regions possessing a rapid oscillation of curvature) of the sealed AAA.
Because of their reduced thickness, the endografts were modelled by means of quadrilateral shell elements.
Reduced integration technique was employed for hexahedral and quadrilateral shell elements.
Since all the components of the sealed AAA consist of incompressible or nearly incompressible material, hybrid formulation for all the finite elements was employed.
Blood was modelled by means of hexahedral shaped solid elements for the static analysis, while hexahedral acoustic elements were employed for the natural frequency analysis; the given blood pressure was exerted on the interior surface of the endografts.
Large displacement, non-linear contact formulation for contact interaction was employed by means of a penalization method allowing for isotropic friction/frictionless sliding and separation of the parts.
Finite elements with linear geometric order and isotropic frictional contact were employed for the static analysis, whereas quadratic geometric order and frictionless contact were employed for the natural frequency analysis.
The idealised model was composed of~\num{240500} nodes and~\num{199603} elements for the static analysis, whereas~\num{382496} nodes and~\num{333579} elements were employed for the frequency analysis.
The model for Patient~1 was composed of~\num{198817} nodes and~\num{473611} elements for the static analysis, whereas~\num{453220} nodes and~\num{622629} elements were employed for the frequency analysis.
The model for Patient~2 was composed of~\num{221250} nodes and~\num{576810} elements for the static analysis, whereas~\num{807599} nodes and~\num{722878} elements were employed for the frequency analysis.

\paragraph{Frequency analysis.}
Acoustic approximation for the linearised Navier-Stokes equation for an incompressible inviscid fluid in laminar flow was employed, thus allowing for a simplification of equations~\eqref{eq:moto} and, contemporarily, maintaining a good approximation of the results~\cite{Carta-Movchan-Argani-Bursi:2016}.
Natural frequency extraction was limited to the interval from~0 to~\SI{30}{\hertz} for Patient~1 and~2, whereas for the idealised model the limit only the first seven natural frequencies were considered (attention was focussed on the range between~0 and~\SI{20}{\hertz}).

\paragraph{Static analysis.}
The blood was modelled as an incompressible medium and its effect is limited to the overall gravity force acting on the system, whereas the blood pressure was modelled as a static constant pressure equal to~\SI{20}{\kilo\pascal} ($ \approx \SI{150}{\mmHg} $) applied directly to the endograft inner surfaces; blood pressure tends to straighten the endografts because they are not rectilinear, which means an additional force that may promote migration.
Three types of static analysis were performed because gravity was applied in the three directions of the reference system ($ x $, $ y $, and $ -z $, see Fig.~\ref{fig:Figura-1}b), representing, respectively, the cases of a patient lying on the back, on a side, or standing vertically.

\section[Acknowledgements]{Acknowledgements}
  \label{Ringraziameti-Informazioni}

The financial support of the Liverpool EPSRC Centre for New Mathematical Sciences Capabilities for Healthcare Technologies, Grant EP/N014499/1 is gratefully acknowledged.

%
%

\phantomsection                          

\end{document}